\documentclass[english, reprint, citeautoscript, groupedaddress, superscriptaddress, 10pt] {revtex4-2}

\usepackage[T1]{fontenc}
\usepackage[latin9]{inputenc}
\usepackage[Symbol]{upgreek}
\usepackage{tikz}
\usepackage[colorlinks=true, linkcolor=black, citecolor=blue, urlcolor=blue, unicode=true]{hyperref}


\begin{document}
\title{Non-reciprocal spin excitations across the skyrmion-paramagnetic phase transition in MnSi}

\newcommand{\cso}{Cu$_{2}$OSeO$_{3}$}

\newcommand{\juelichill}{J\"ulich Centre for Neutron Science (JCNS), Outstation at ILL, 71 avenue des Martyrs, 38000 Grenoble, France}
\newcommand{\cologne}{Institut f\"ur Theoretische Physik, Universit\"at zu K\"oln, Z\"ulpicher Str. 77a, 50937 K\"oln, Germany}
\newcommand{\kit}{Karlsruhe Institute of Technology (KIT), 76131 Karlsruhe, Germany}
\newcommand{\ill}{Institut Laue-Langevin (ILL), 71 avenue des Martyrs, 38000 Grenoble, France}
\newcommand{\tum}{Physik-Department, Technische Universit\"at M\"unchen (TUM), James-Franck-Str. 1, 85748 Garching, Germany}
\newcommand{\frm}{Forschungsneutronenquelle Heinz-Maier-Leibnitz (MLZ), Lichtenbergstr. 1, 85747 Garching, Germany}
\newcommand{\mcqst}{Munich Center for Quantum Science and Technology (MCQST), Schellingstr. 4, 80799 Munich, Germany}
\newcommand{\zqe}{Zentrum f\"ur QuantumEngineering (ZQE), Am Coulombwall 3a, 85748 Garching, Germany}

\author{T. Weber}
\email[Corresponding author: ]{tobias.weber@ill.fr}
\affiliation{\ill}

\author{K. Schmalzl}
\affiliation{\juelichill}

\author{J. Waizner}
\affiliation{\cologne}

\author{A. Bauer}
\affiliation{\tum}

\author{M. Garst}
\affiliation{\kit}

\author{C. Pfleiderer}
\affiliation{\tum}
\affiliation{\frm}
\affiliation{\mcqst}
\affiliation{\zqe}

\date{\today}


\begin{abstract}
The magnetic excitations of the skyrmion lattice in MnSi comprise a multitude of individual modes,
which are non-reciprocal and thereby propagate unidirectionally.
We report inelastic neutron scattering experiments for temperatures near and above the skyrmion-paramagnetic
phase transition in the chiral magnet MnSi tracking the evolution from the
skyrmion lattice towards the high-temperature paramagnetic state.
Within the resolution of the triple-axis measurements the
excitations vary smoothly across the skyrmion-paramagnetic boundary, and,
the quasi-elastic paramagnetic signal under applied field
retains the non-reciprocal character seen
in the skyrmion phase even far above the critical temperature.
Using a resolution-convolution our results are consistent with linear spin-wave theory.
\end{abstract}

\keywords{magnons; skyrmion lattice; inelastic neutron scattering; linear spin-wave theory}

\maketitle
\section{Introduction}
Below temperatures of $T_c \approx 29\, \mathrm{K}$,
the itinerant-electron compound MnSi features several magnetically ordered phases \cite{Bauer2012, Bauer2013},
which have drawn great interest in the last decade.
The magnetic phase diagram comprises a helical, conical, field-polarised ferromagnetic,
as well as a skyrmion state which is topologically distinct in terms of its non-zero winding number \cite{Muehl2009}.
In the helical phase below a critical field $B_{c1} \approx 100\ \mathrm{mT}$,
four helical domains align along the $\left[111\right]$ easy axes \cite{Ishikawa1976}.
Increasing the field above $B_{c1}$ aligns the multiple domains into one single domain,
with the helix propagation vector ordering along the applied field direction \cite{Kugler2015, Muehl2009}.
Further increasing the field causes a conical canting of the spins towards the field direction until,
above the second critical field of $B_{c2} \approx 600\ \mathrm{mT}$,
they are fully aligned with the external field.

Spontaneous long-range oder vanishes above a transition temperature of $T_c \approx 29\,\mathrm{K}$ \cite{Ishikawa1985, Roessli2002}.
A fluctuation-disordered paramagnetic regime exists between the long-range ordered phases and the paramagnetic regime
at high temperatures in which the discrete magnetic satellites of the ordered phases spread out evenly over the surface of a sphere
centred around the nuclear Bragg reflections \cite{Jano2013}.

Studies of the spin excitations of the ordered phases of MnSi go back into the 1970s
with the pioneering work by the Ishikawa group \cite{Ishikawa1977}.
They could decipher the general parabolic form the
the dispersion, but their experiments were not sensitive to the details of the dispersion branches.
In later high-resolution measurements, magnetic excitations in the helical phase were found to take the
form of a band structure for momentum transfers
$q$ perpendicular to the helix propagation vector and of two non-reciprocal
as well as one central, symmetric mode for momenta along the helix \cite{Jano2010Heli, Kugler2015, Weber2018a, Weber2019Coni}.
The magnon band structure of the helical and conical phases originates from a strong back-folding of the
spectra into the first magnetic Brillouin zone \cite{Garst2017},
which -- for MnSi -- is approximately 35 times smaller than the first nuclear zone.

Non-reciprocity in the simplest scenario alludes to magnon dispersions,
where magnons are created at different magnitudes of energy than they are annihilated \cite{Sato2019}.
It may be the consequence of a non-centrosymmetric crystal structure -- MnSi crystallises in the $\mathrm{P2_13}$ space group --
and the ensuing Dzyaloshinskii-Moriya interaction
together with the broken time-reversal symmetry due to a spin order that is, in the present case, imposed by an external magnetic field \cite{Sato2019}.
The investigation of non-reciprocal magnon dispersions holds a great potential in the research of magnonic devices where the propagation
of magnons in one direction is required, such as spin-wave diodes \cite{Szulc2020} or directional couplers \cite{Tian2025}.

The asymmetry of the non-reciprocal modes in MnSi becomes more pronounced in the conical phase, with the spectral weight shifting
from one of the modes centred on a magnetic satellite to the other, until only a single mode remains in the fully field-polarised phase,
This phenomenon was first mentioned in the 1980s \cite{Shirane1983}.
Interestingly, the mode in the field-polarised ferromagnetic state rests centred on a position in momentum space
where one of the magnetic satellites of the conical phase would be,
even though the field-polarised phase in commensurate and thus no satellite peaks remain \cite{Weber2018a}.

First investigations into the excitations of the paramagnetic phase were conducted in the 1980s \cite{Ishikawa1985}.
Asymmetric, polarisation-dependent fluctuations could be identified to persist beyond $T_c$ \cite{Roessli2002, Roessli2004}.

In the skyrmion phase, a multitude of non-reciprocal modes are excited for momentum transfers perpendicular
to the skyrmion plane.
For momentum transfers inside the skyrmion plane, on the other hand,
the magnon modes back-fold into the first magnetic Brillouin zone and
create complicated Landau levels, similar to the band formation in the helical phase \cite{Jano2010Skyrmi, Weber2022Skyrmi}.
In a recent work, Soda \textit{et al.} could confirm that the asymmetry of the non-reciprocal modes extends to the
$\mathrm{\upmu eV}$ region \cite{Soda2023}.

An investigation into the evolution of skyrmion dispersion inside the first magnetic Brillouin
zone of the helimagnet \cso{}, which shares many properties of MnSi,
towards the field-polarised regime was published very recently \cite{Che2024}:
Using Brillouin light scattering, they could show how the individual modes of the skyrmion lattice merge into a
single field-aligned ferromagnetic mode.
For MnSi we previously investigated the transitions from the conical phase \cite{Weber2018a}
and found a qualitatively similar picture for the transition towards the ferromagnetic dispersion.

First studies on the magnetisation dynamics at the skyrmion-paramagnetic transition were performed by
Schwarze \textit{et al.} and Kindervater \textit{et al.}, where they could discern the changes of the modes from the different phases
using microwave techniques \cite{Schwarze2015, Kindervater2019}.
The microwave technique that both groups employed operates at zero momentum transfer, $q = 0$,
and is thus restricted to probing the centre of the nuclear Brillouin zone.
In this specific regime, they could attribute the fundamental excitations of the skyrmion lattice to counter-clockwise and
clockwise rotations as well as a breathing motion of the skyrmions.

For our present work we investigated the spin excitations in MnSi close to the transitions from the skyrmion
phase under increasing temperatures.
Our goal was to test the stability and evolution of the magnons of the skyrmion phase and whether we could
identify a clear separation of the excitations across the phases.
We employed inelastic neutron scattering which does not restrict us to zero momentum transfer.

\section{Skyrmion-paramagnetic transition}
\label{sec:IN12_para}
\subsection{Overview and Experimental Setup}
As part of a larger investigation into the stability of the magnon modes at the border of the skyrmion phase,
the principal part of the present experiment focused on the skyrmion-paramagnetic transition.
The experiment was conducted at the cold-neutron triple-axis spectrometer \textit{IN12} \cite{IN12} at the ILL.
We used horizontal collimations of 30 minutes both before and after the MnSi sample.
The instrument possesses a velocity selector in the neutron guide before the monochromator crystals.
In addition, a cooled beryllium crystal was placed in the instrument's $k_f$ axis between the MnSi sample
and the instrument's analyser crystals.
The beryllium crystal suppresses residual higher energy neutrons not completely removed by the velocity selector,
prevents spurious higher order contamination of the analyser, and reduces the instrument's background.
For the experiment, we chose fixed final wavenumbers in the range of $k_f = 1.4 - 1.5 \textup{~\AA}^{-1}$.

\begin{figure}[htb]
\begin{centering}
	\includegraphics[width=0.75\columnwidth]{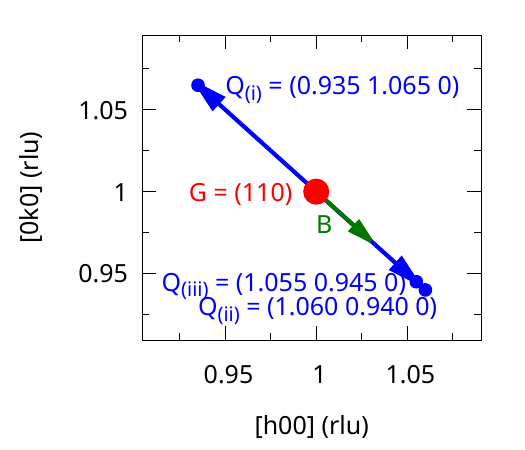}
	\caption{Experimental set-up showing the two principal momentum transfers, $Q_{(i)}$ and $Q_{(ii)}$.
	We measured in the $\left<hk0\right>$ scattering plane
	at total momentum transfers $Q = G + q$ and external field $B$ transverse to the $G = \left( 110 \right)$ Bragg peak.}
	\label{fig:scanpos_para}
\end{centering}
\end{figure}

We used the same single crystal sample as in our previous studies \cite{Weber2022Skyrmi, Weber2019Coni, Weber2018b, Weber2018a}.
The crystal is of cylindrical geometry, with a diameter of one cm and a height of three cm, and has a mass of about 15 g,
with the cylinder's long axis pointing along a $\left[001\right]$ direction.
The sample was mounted in a horizontal \textit{Oxford} magnet \cite{MagnetH}
with the $\left[100\right]$ and $\left[010\right]$ directions in the horizontal scattering plane.
The magnetic field was set to $B = \pm 195\,\mathrm{mT}$ along $\left[1\bar{1}0\right]$.
The given value for $B$ corresponds to the magnitude of the externally applied field,
whereas for the calculations we also take into account the demagnetisation factor due to the sample geometry \cite{Sato1989}.
The scattering geometry around the $\left( 110 \right)$ Bragg peak is depicted in Fig. \ref{fig:scanpos_para}.
There, the principal scan positions are marked as $Q_{(i)}$ and $Q_{(ii)}$,
where $Q$ names the total momentum transfer as the sum of the reciprocal lattice vector
and the reduced momentum transfer, $Q = G + q$.

\begin{figure*}[htb]
\begin{centering}
	\begin{tikzpicture}
		\draw (0, 0) node [ inner sep = 0 ] { \includegraphics[width=0.49\textwidth]{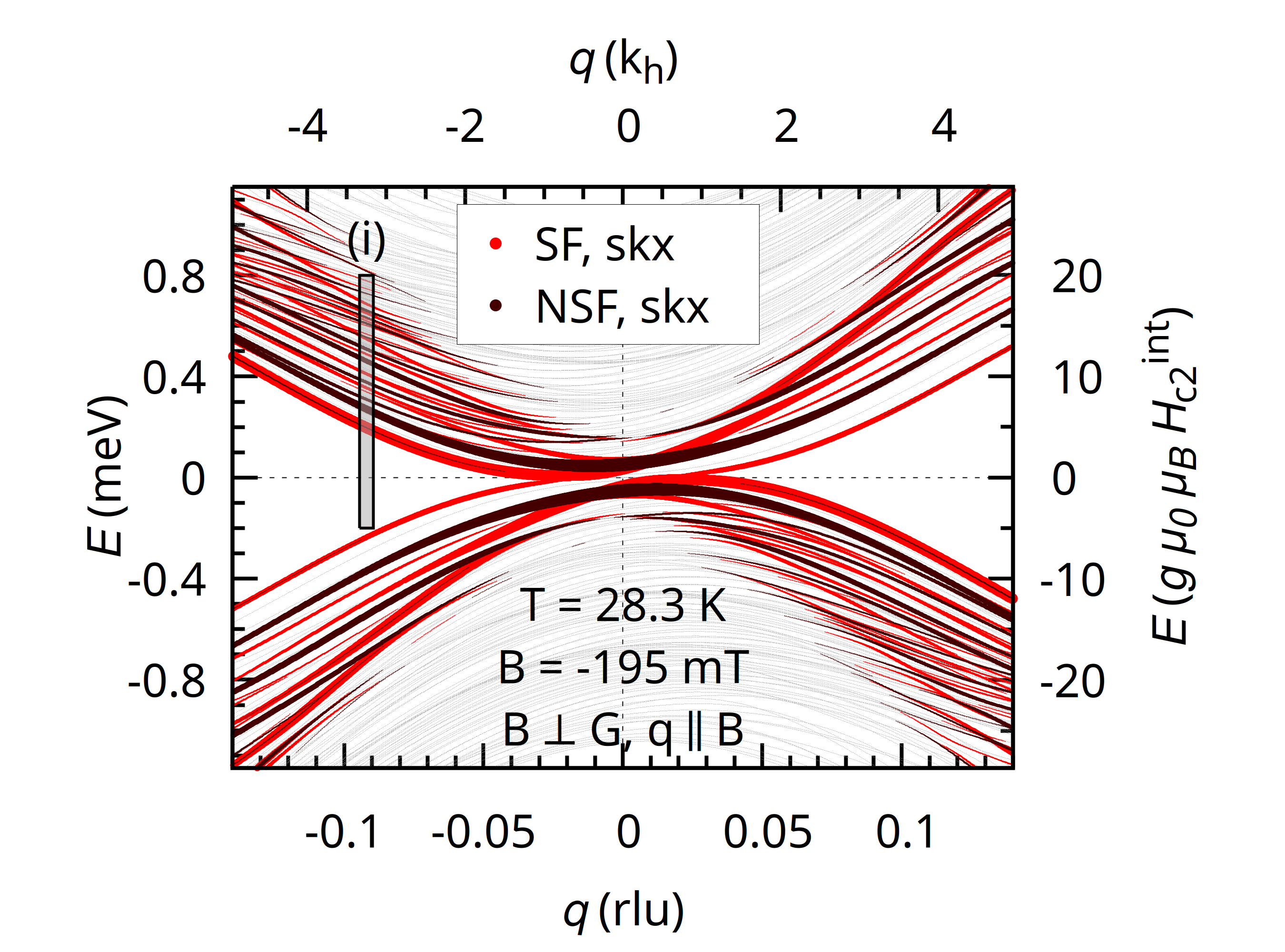} };
		\draw (-3.5, 2.25) node { \bf \fontfamily{phv} \selectfont (a) };
	\end{tikzpicture}
	\begin{tikzpicture}
		\draw (0, 0) node [ inner sep = 0 ] { \includegraphics[width=0.49\textwidth]{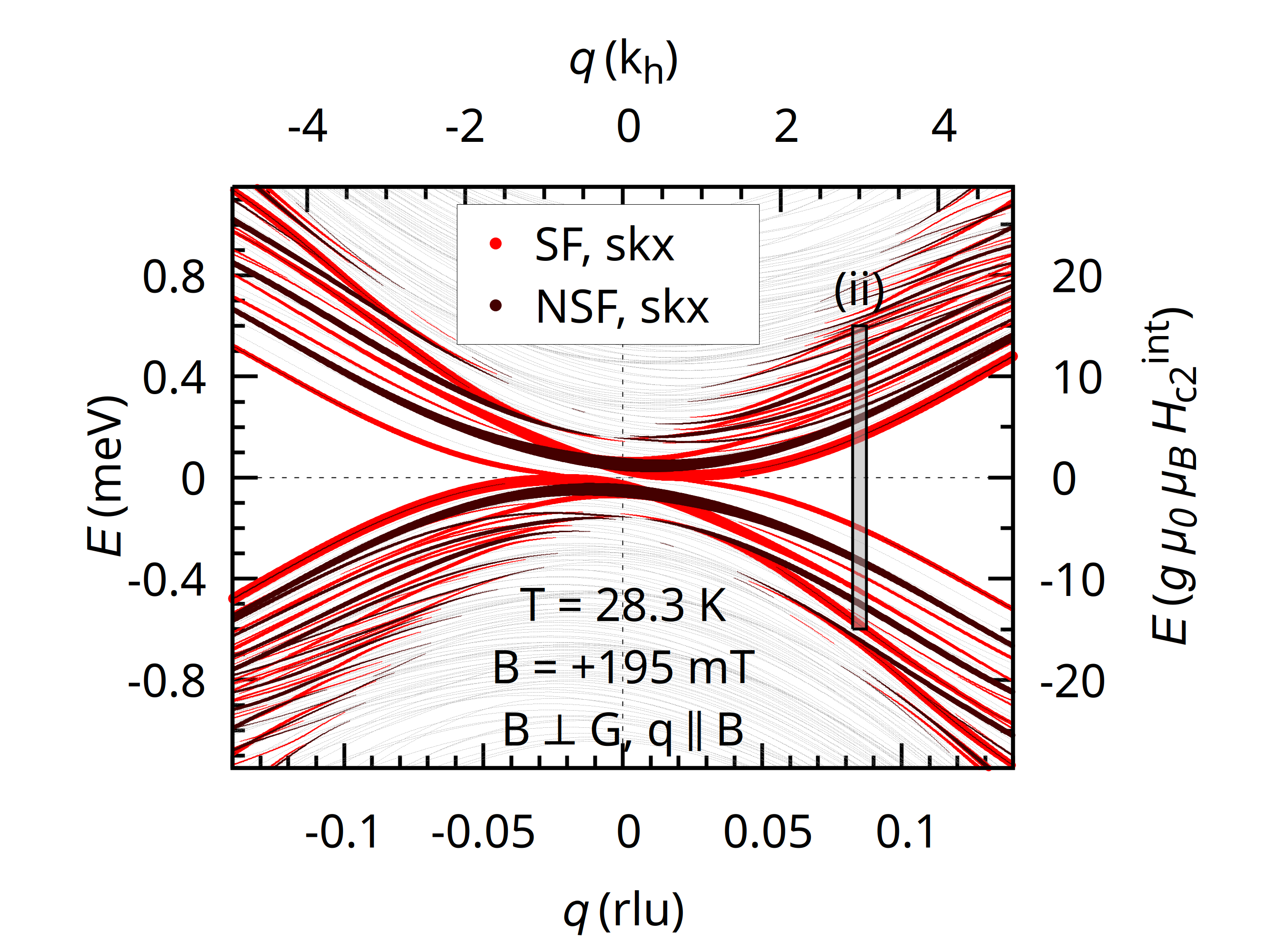} };
		\draw (-3.5, 2.25) node { \bf \fontfamily{phv} \selectfont (b) };
	\end{tikzpicture}
	\caption{Calculations using our previously developed linear
	spin-wave model \protect\cite{Garst2017, Waizner2016phd, Weber2022Skyrmi}
	for the skyrmion order at $28.3\,\mathrm{K}$.
	The plots show the magnon modes propagating along $\left(110\right) + q \cdot \left[1\bar{1}0\right]$.
	Spin-flip (SF) and non-spin-flip (NSF) components of the scattering cross sections are depicted as red and black curves, respectively.
	The experiment itself was unpolarised, summing the NSF and SF channels.
	The thickness of the lines symbolises the spectral weights of the modes.
	The grey bars labelled (i) and (ii) mark the positions of the scans in Fig. \ref{fig:IN12_skx_T}.
	Panels (a) and (b) depict the dispersion for $B = -195\ \mathrm{mT}$ and $B = +195\ \mathrm{mT}$, respectively.}
	\label{fig:theo_skx}
\end{centering}
\end{figure*}

Fig. \ref{fig:theo_skx} depicts the results of a theoretical calculation
based on our previously developed model \cite{Garst2017, Waizner2016phd, Weber2022Skyrmi}
of the magnon dispersion in the skyrmion phase.
It is dominated by energetically closely-spaced parabolic magnon modes.
The theoretical model is based on the Landau-Lifshitz equation and takes into account
symmetric exchange, Dzyaloshinskii-Moriya and Zeeman terms \cite{Waizner2016phd},
as well as a higher-order correction in the gradient expansion \cite{Kugler2015}.
The non-reciprocal nature of the excitations is visible by the off-centring of the branches.
Both panels of Fig. \ref{fig:theo_skx} show the same section of momentum transfer,
with only the direction of the field inverted in panel (b) with respect to (a).
The thickness of the lines corresponds to the spin-spin correlation function,
which yields the spectral weight of a mode.
The abscissas of the plots show the reduced momentum transfer both in reciprocal lattice units, rlu, and in helix wavenumbers,
$\mathrm{k_h} \approx 0.039 \textup{~\AA}^{-1} \approx 0.028\ \mathrm{rlu}$.
The ordinates show the energy transfer in meV and in units relative to the energy at the conical-field polarised
phase transition, $g \mu_0 \mu_B H_{c2}^{int}$,
with $g \approx 2$ for electrons and $H_{c2}^{int}$ the sample's internal field at the transition,
including demagnetisation effects \cite{Sato1989}.
The centre of the figure corresponds to $q = 0$,
where the clockwise, counter-clockwise, and breathing modes of the skyrmion lattice had been discerned previously \cite{Schwarze2015, Kindervater2019}.
At finite momentum transfer, $q \neq 0$, a multitude of modes appear.

For the experiment we chose momentum transfer vectors, $q_{(i)}$ and $q_{(ii)}$ where the magnon
modes of the skyrmion phase are clearly visible, these are shown as grey vertical bars labelled (i) and (ii) in Fig. \ref{fig:theo_skx}.
In the present experiment, we repeated the same scans that we performed in the skyrmion phase for increasing temperatures
where the skyrmion lattice vanishes and gives way to paramagnetism.

\subsection{Results}
\begin{figure*}[htb]
\begin{centering}
	\begin{tikzpicture}
		\draw (0, 0) node [ inner sep = 0 ] { \includegraphics[width=0.44\textwidth]{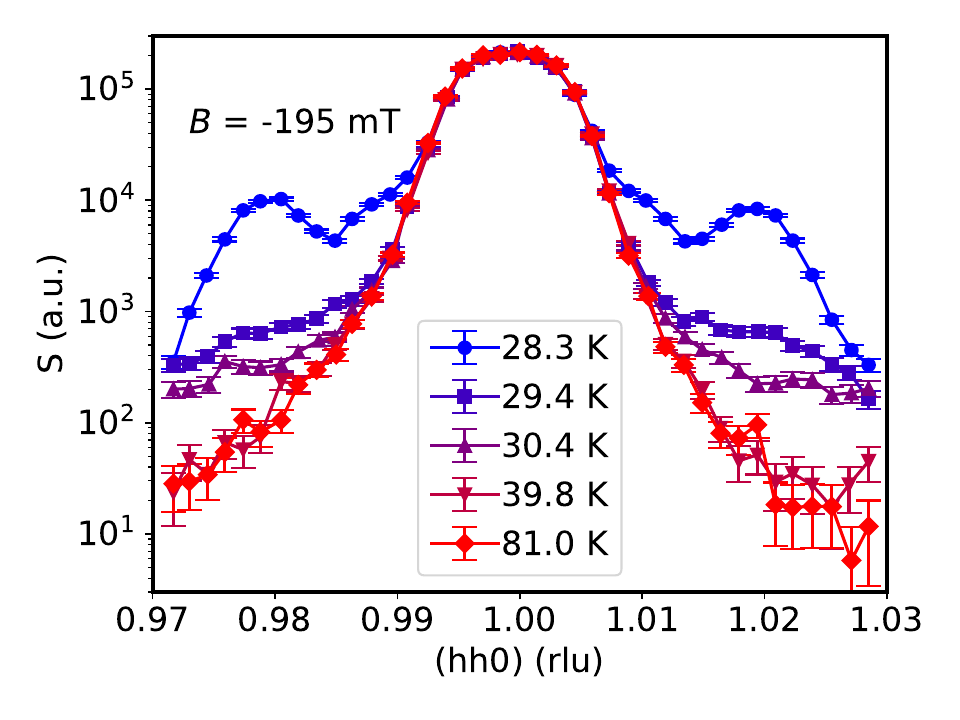} };
		\draw (-3.4, 1.9) node { \bf \fontfamily{phv} \selectfont (a) };
	\end{tikzpicture}
	\begin{tikzpicture}
		\draw (0, 0) node [ inner sep = 0 ] { \includegraphics[width=0.44\textwidth]{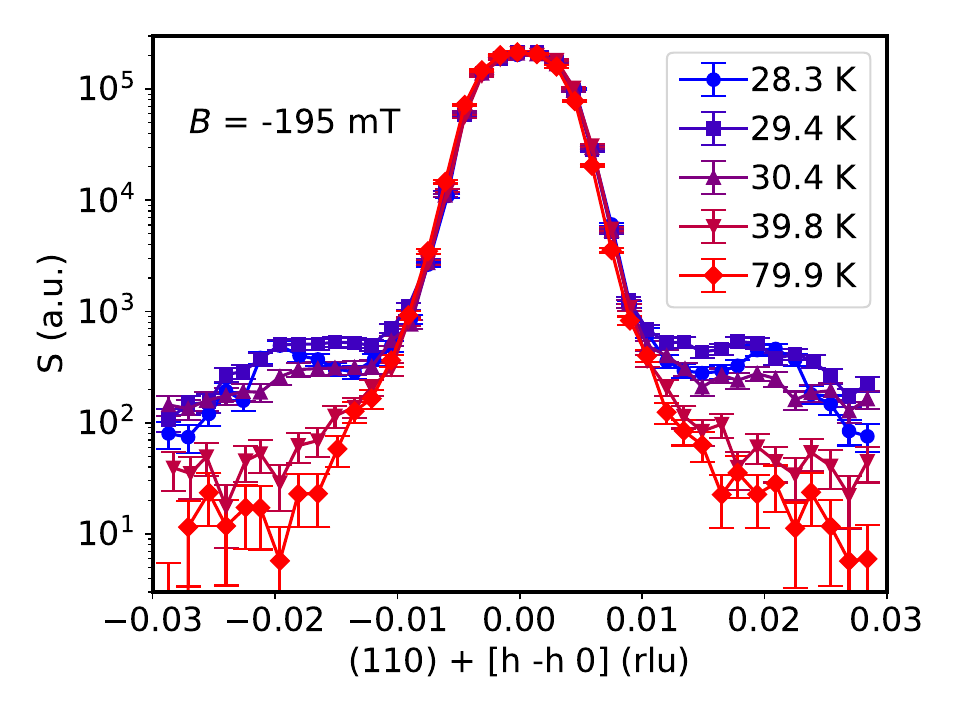} };
		\draw (-3.4, 1.9) node { \bf \fontfamily{phv} \selectfont (b) };
	\end{tikzpicture}

	\begin{tikzpicture}
		\draw (0, 0) node [ inner sep = 0 ] { \includegraphics[width=0.42\textwidth]{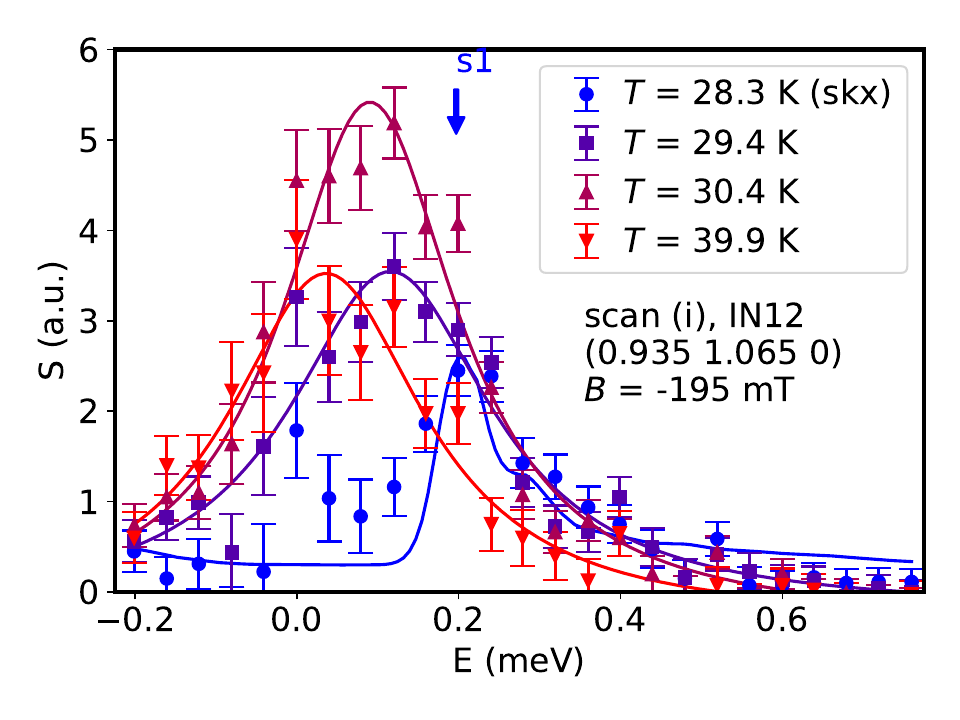} };
		\draw (-3.4, 1.9) node { \bf \fontfamily{phv} \selectfont (c) };
	\end{tikzpicture}
	\begin{tikzpicture}
		\draw (0, 0) node [ inner sep = 0 ] { \includegraphics[width=0.42\textwidth]{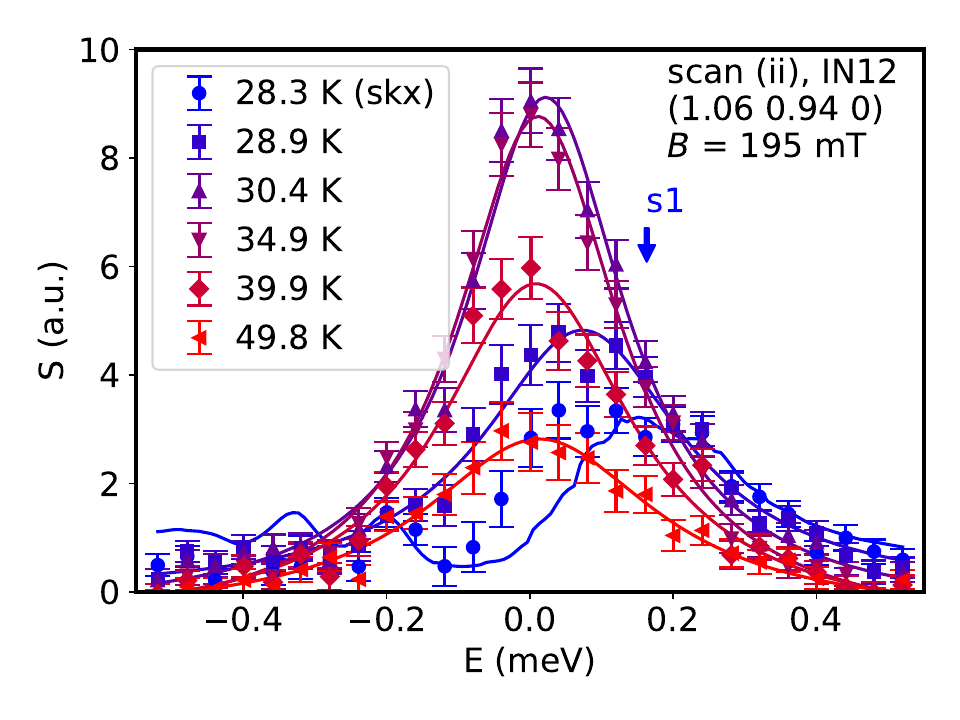} };
		\draw (-3.4, 1.9) node { \bf \fontfamily{phv} \selectfont (d) };
	\end{tikzpicture}

	\begin{tikzpicture}
		\draw (0, 0) node [ inner sep = 0 ] { \includegraphics[width=0.44\textwidth]{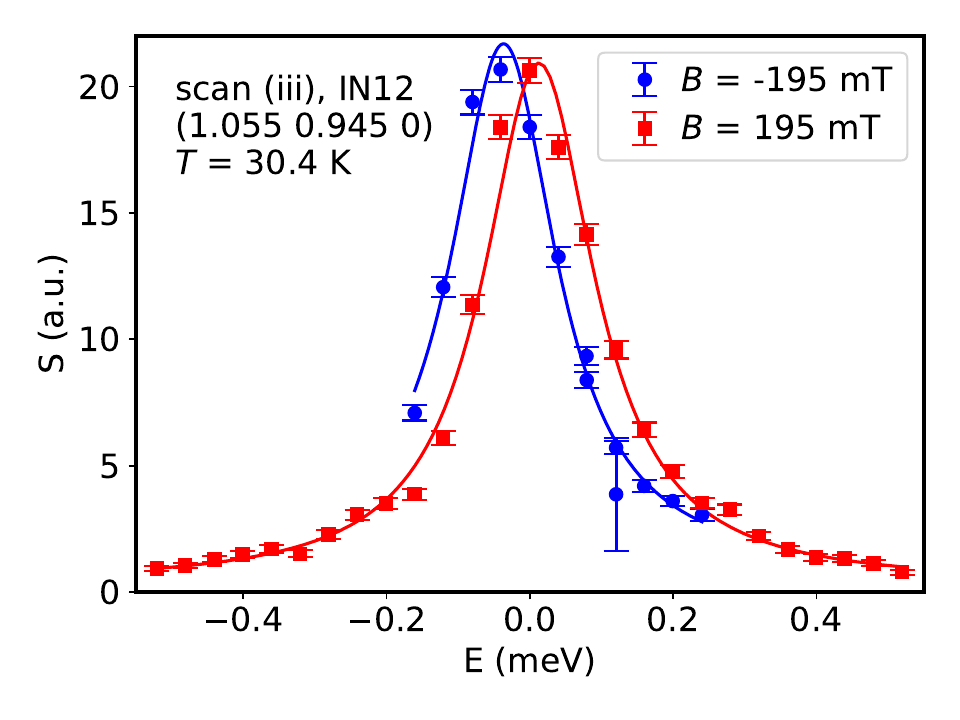} };
		\draw (-3.4, 1.9) node { \bf \fontfamily{phv} \selectfont (e) };
	\end{tikzpicture}
	\begin{tikzpicture}
		\draw (0, 0) node [ inner sep = 0 ] { \includegraphics[width=0.44\textwidth]{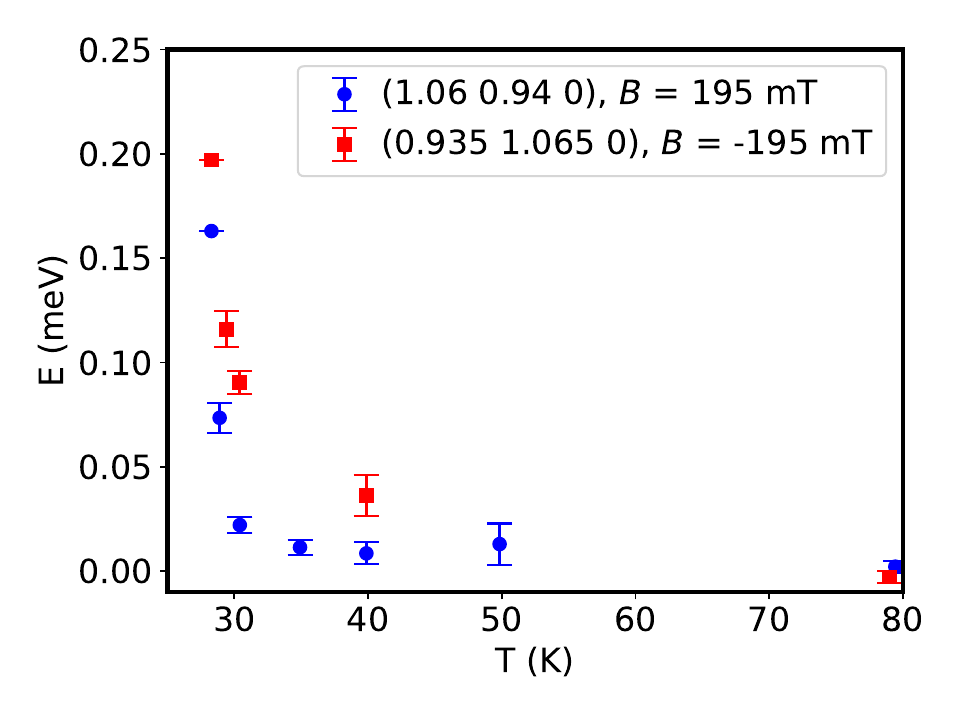} };
		\draw (-3.4, 1.9) node { \bf \fontfamily{phv} \selectfont (f) };
	\end{tikzpicture}

	\caption{(a) Longitudinal elastic scan around $\left( 110 \right)$. (b) Transversal elastic scan around $\left( 110 \right)$.
		(c) and (d): Inelastic scans showing the temperature-dependent evolution of the magnon modes starting from the skyrmion phase
		at $T=28.3\,\mathrm{K}$ and up to the nonmagnetic phase.
		The scans in panels (c) and (d) correspond to respective positions (i) and (ii) marked in Fig. \ref{fig:theo_skx}.
		The complicated magnon structure in the skyrmion phase is lost when increasing the temperature.
		The solid lines for the skyrmion phase at $T_{skx}=28.3\,\mathrm{K}$ are resolution-convolution
		simulations of the magnon model \protect\cite{Garst2017, Weber2022Skyrmi},
		all the other solid lines are Lorentzian fits.
		The label \textit{s1} marks the position where the first excitation of the skyrmion lattice is expected.
		(e) and (f): The non-reciprocity that is characteristic of the ordered magnetic phases is retained throughout
		the paramagnetic phase and only disappears in the clearly non-magnetic regime for $T>80\,\mathrm{K}$.
		It manifests itself via a time-reversal asymmetry that is evident when flipping the polarity of the magnetic field.
		Panel (e) shows the paramagnons for $T=30.4\,\mathrm{K}$. Here, the solid lines are Lorentzian fits.
		In panel (f) the asymmetric energy maxima are plotted against temperature.}
	\label{fig:IN12_skx_T}
\end{centering}
\end{figure*}
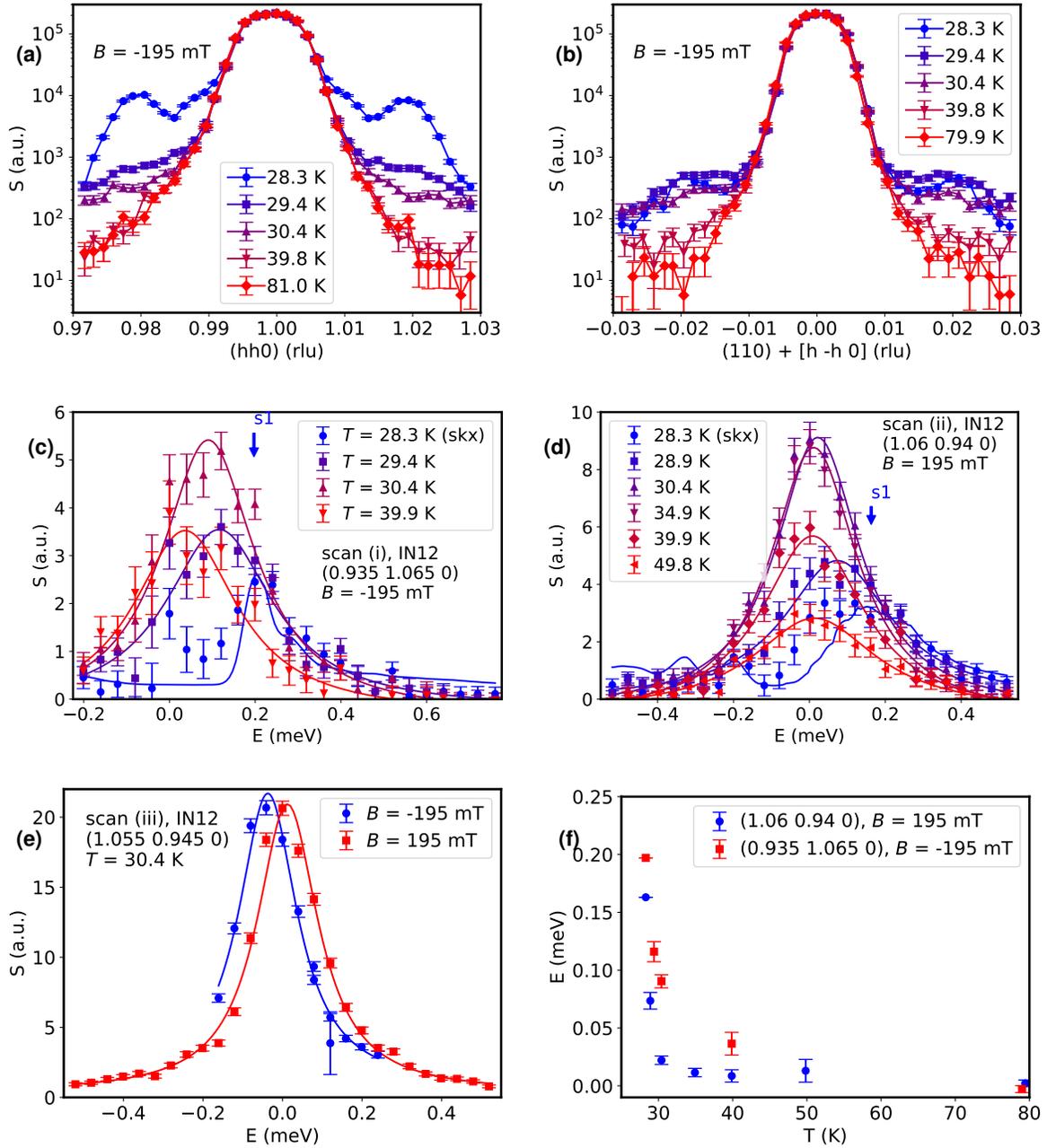

\paragraph{Elastic scattering.}
As first step we determined the exact temperatures of the phase boundaries.
Panels (a) and (b) of Fig. \ref{fig:IN12_skx_T} show longitudinal and transverse elastic scans around the $\left(110\right)$
nuclear Bragg peak.
Two skyrmion satellite reflections are clearly visible in panel (a) at $T = 28.3\,\mathrm{K}$ and $q = \pm0.02$,
the projections onto the scattering plane of the other four peaks can be discerned at $q = \pm0.01$.
The projections originate from four skyrmion peaks that are above and below the $\left<hk0\right>$ plane,
but are observable in-plane due to the instrument's resolution.
The peaks disappear towards higher temperatures.
Between $T = 29.4\,\mathrm{K}$ and $T = 30.4\,\mathrm{K}$, the spherically spread elastic signal \cite{Jano2013}
of the fluctuation-disordered phase is visible.
For still higher temperatures, all magnetic satellites disappear.

\paragraph{Inelastic scattering.}
In all of the previously identified phases we performed inelastic neutron scattering as the second step.
Fig. \ref{fig:IN12_skx_T} (c) depicts how the magnon modes of the skyrmion lattice
at $Q_{(i)} = \left( 0.935,\; 1.065,\; 0 \right)$ and $T = 28.3\,\mathrm{K}$ merge into paramagnetic excitations
($T = 29.4 - 39.9\,\mathrm{K}$) and finally vanish for high temperatures ($T = 79\,\mathrm{K}$).
Purely non-magnetic data collected at $T = 79\,\mathrm{K}$ was subtracted from all other data sets.
The full non-subtracted data sets are shown in Appendix \ref{sec:data_appendix}.

The curve shown for the $T = 28.3\,\mathrm{K}$ data in Fig. \ref{fig:IN12_skx_T} (c) is a Monte-Carlo
resolution-convolution \cite{Popovici1975, Takin2023} of the instrumental
resolution and the theoretical linear spin-wave model \cite{Garst2017, Weber2022Skyrmi} that we described in the previous section,
all other curves are simple Lorentzian fits, which have been found to describe the paramagnons well.
The spin-wave model for the skyrmion phase itself is parameter-free,
the only free variable for the resolution-convolution
was a global scaling parameter for the dynamical structure factor.

The modes of the skyrmion lattice cannot be discerned individually as they are far below the resolution
limit of any triple-axis spectrometer, but they can be well reproduced via the convolution of the theory,
where they appear as broad bands in the spectrum.
The results show that upon leaving the skyrmion phase by heating the sample up,
the little signatures discernible in the skyrmion phase broadens and merges into featureless nearly quasi-elastic spectra.

In Fig. \ref{fig:IN12_skx_T} (d) the $Q_{(ii)} = \left( 1.06,\; 0.94,\; 0 \right)$ magnon modes in the skyrmion phase are visible,
but are not as pronounced as in the previous measurement,
because here positive energy transfer corresponds to transverse defocusing of the instrument.
This served as a check against possible spurions, namely Bragg tails \cite{Shirane2002}, in the focusing scans.
Bragg tails are remnants of strong Bragg peaks that appear as false inelastic signals due to the correlation of momentum and energy in the instrumental resolution function \cite{Popovici1975}.
In the skyrmion phase, this problem is especially severe as each of the magnetic satellite reflections generates a Bragg tail in addition to the nuclear peak.
The polarity of the magnetic field is flipped for the scans at momentum transfer $Q_{(ii)}$
with respect to the setup used for $Q_{(i)}$.
As we simultaneously invert the direction of the reduced momentum, $q_{(ii)} = Q_{(ii)} - G_{\left(110\right)}$,
in comparison to $q_{(i)} = Q_{(i)} - G_{\left(110\right)}$ the physics does not change.
Non-reciprocity implies that the original dispersion and dynamical structure factor is recovered
when changing the signs of both field and reduced momentum.

As before, the $T = 28.3\,\mathrm{K}$ curve is a resolution-convolution of the model \cite{Garst2017, Weber2022Skyrmi},
all other curves are Lorentzian fits. High-$T$ data has been subtracted.
As the paramagnons move to lower energies for increasing temperatures,
their intensities increase due to the Bose factor.
Their intensities observed around $E = 0$ reach a maximum at approximately $T = 35\,\mathrm{K}$,
for even higher temperatures they decrease again until only nuclear-incoherent scattering is left at $T > 79\,\mathrm{K}$.

The non-reciprocal character of the magnetic excitations, that is found in all of the ordered phases, is retained in the paramagnetic regime,
but becomes less pronounced as the temperatures approach the non-magnetic regime.
Fig. \ref{fig:IN12_skx_T} (e) shows a scan at $Q_{(iii)} = \left( 1.055,\; 0.945,\; 0 \right)$
in the paramagnetic phase at $30.4\ \mathrm{K}$ for two directions of the external field.
This visualises the time reversal asymmetry that is observed upon inverting the direction of the magnetic field.
The same dispersion would only be recovered when flipping both the reduced momentum transfer
and the magnetic field direction at the same time.
Please note that within the instrumental resolution the points $Q_{(iii)}$ and $Q_{(ii)}$ and their
dispersions are virtually the same (see Fig. \ref{fig:scanpos_para}), their distinction only exists for technical reasons.

Due to non-reciprocity, the spin excitations are not centred around $E=0$,
even though the system is deep in its paramagnetic state.
The temperature-dependent energy shifts of the maxima of the scattering intensity is plotted in Fig. \ref{fig:IN12_skx_T} (f)
for the two principal scan positions, $Q_{(i)}$ and $Q_{(ii)}$.
For the data outside the skyrmion phase,
the energy shifts were obtained from the centres of the Lorentzian fits shown in panels (c) and (d).
For the data inside the skyrmion phase, we calculated the energies of the modes using our theoretical model.
At temperatures as high as $49.8\ \mathrm{K}$ an energy offset can still be discerned.

\clearpage

\section{Helimagnetic-paramagnetic transition}
\label{sec:IN12_para_heli}
The final part of the experiment concerned the helimagnetic-paramagnetic transition for no applied external field.
A large MnSi single crystal of approximately 50 g was oriented in a $\left<hk0\right>$ scattering plane
and placed in a standard ILL \textit{Orange} cryostat.
As in the first part, the experiment was conducted using horizontal collimators of 30 minutes both before and after the MnSi sample
and with a beryllium crystal in the instrument's $k_f$ axis.
We used a fixed $k_f$ of $1.5 \textup{~\AA}^{-1}$.

\begin{figure}[htb]
\begin{centering}
	\begin{tikzpicture}
		\draw (0, 0) node [ inner sep = 0 ] { \includegraphics[width=0.44\textwidth]{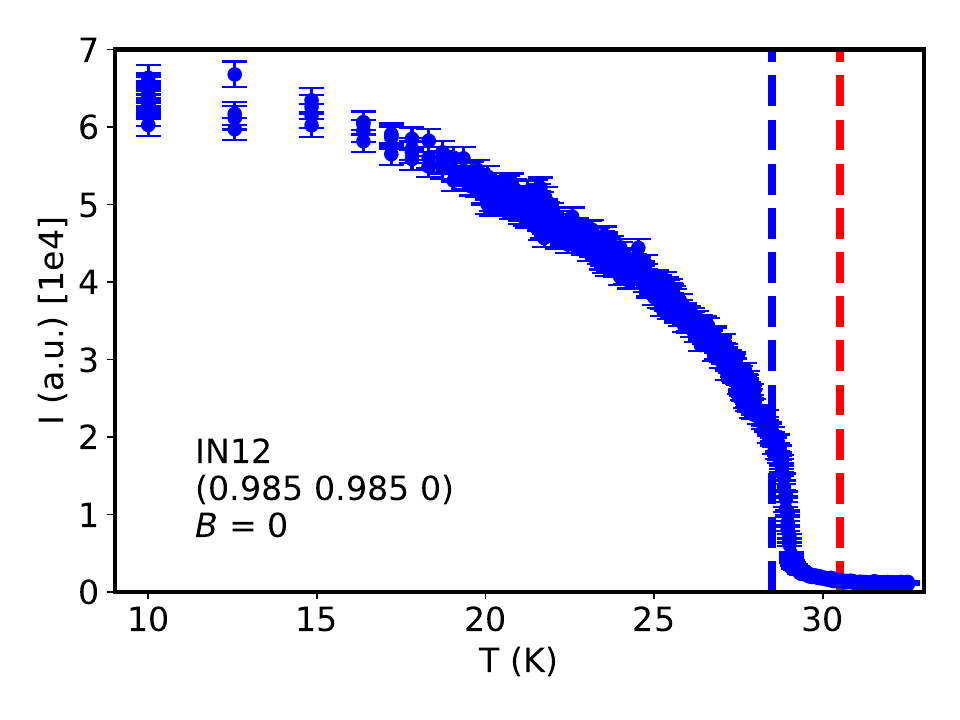} };
		\draw (-3.3, 2.05) node { \bf \fontfamily{phv} \selectfont (a) };
	\end{tikzpicture}
	\begin{tikzpicture}
		\draw (0, 0) node [ inner sep = 0 ] { \includegraphics[width=0.44\textwidth]{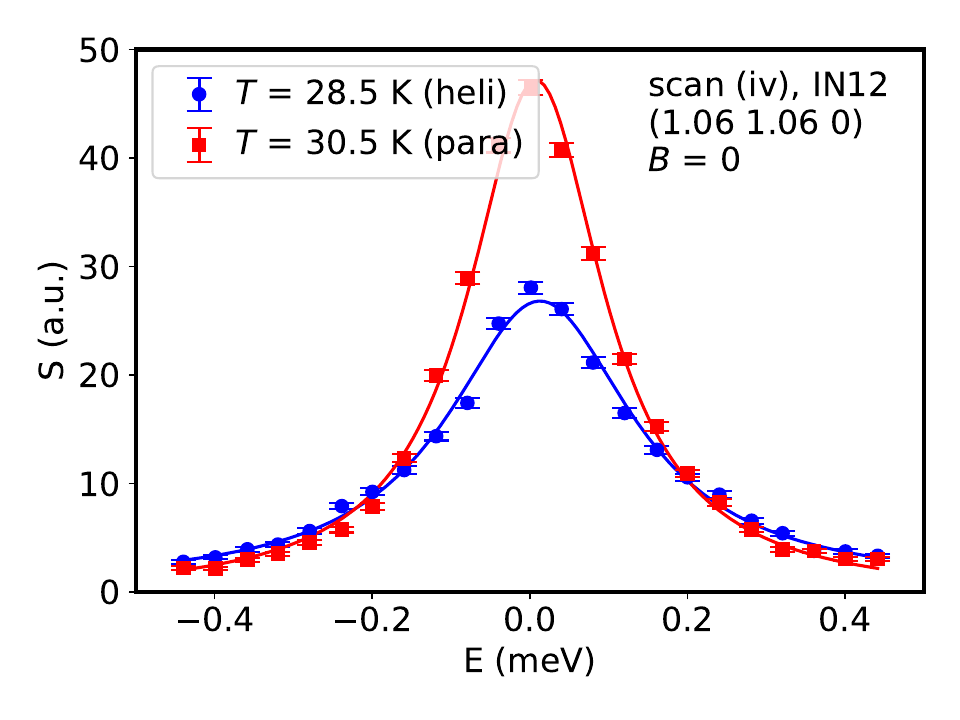} };
		\draw (-3.3, 2.05) node { \bf \fontfamily{phv} \selectfont (b) };
	\end{tikzpicture}
	\caption{(a) Elastic scan of one of the helimagnetic satellites vs. temperature.
	The phase transition towards paramagnetism is observed at $T_c = 29\,\mathrm{K}$.
	The vertical dashed lines mark the scan temperatures.
	(b) Magnon modes in the multi-domain helical ($T=28.5\,\mathrm{K}$)
	and in the paramagnetic ($T=30.5\,\mathrm{K}$) phase at $Q = \left(1.06,\; 1.06,\; 0 \right)$
	close to their phase boundary.
	The solid lines are Lorentzian fits.}
	\label{fig:IN12_heli_T}
\end{centering}
\end{figure}

As no magnetic field was applied, the helimagnetic state comprises four domains with the
helices aligned along the $\left<111\right>$ directions \cite{Kugler2015}.
Magnons emanating from all four domains overlap in this state \cite{Jano2010Heli}.
Fig. \ref{fig:IN12_heli_T} (a) tracks the intensity of a projection onto the $\left[110\right]$ plane of one of these peaks
against temperature. The phase transition towards paramagnetism sets in at $T_c = 29\, \mathrm{K}$.

Fig. \ref{fig:IN12_heli_T} (b) depicts the spin excitations of both the helimagnetic and the paramagnetic state
at $Q = \left(1.06,\; 1.06,\; 0 \right)$ close to their common phase boundary.
The overlap of magnons from four magnetic domains is evidenced by the broad tails around the central incoherent peak
($\Gamma_{\mathrm{FWHM}} = 0.287\, \mathrm{meV} \pm 0.006\, \mathrm{meV}$).
Heating the sample beyond the phase transition shifts the peak of the intensity towards lower energies, which manifests itself by
a higher amplitude but lower width of the observed peak ($\Gamma_{\mathrm{FWHM}} = 0.207\, \mathrm{meV} \pm 0.003\, \mathrm{meV}$).
The values are from Lorentzian curve fits, a resolution-deconvolution or convolution simulation was not attempted.
At zero field, the line shapes are entirely symmetrical, a non-reciprocity is not observed.

\section{Summary and Discussion}
We investigated the transition of the magnon dispersion from the skyrmion to the paramagnetic phase of MnSi at momentum transfers that
are not restricted to the Brillouin zone centre.

Measuring the excitations of the skyrmion phase and beyond proves challenging,
since the energy scales are on the brink of what is resolvable with inelastic neutron scattering.
In the skyrmion phase, the magnons comprise a plethora of individual modes, which, while not being observable individually,
can be well reproduced by convoluting the linear spin-wave model with the instrumental resolution.

While the data alone suggests a smooth transition in temperature upon changing into the paramagnetic state,
we infer from the convolution of the theoretical model that the magnons still experience a clear transition as far
as the internal details of the dispersion are concerned.
Above the transition temperature, the complicated inelastic modes
give way to quasi-elastic broadenings while retaining their non-reciprocal characteristics.
In the fluctuation-disordered phase, a strong increase in intensity can be observed,
which is due to the Bose occupation factor at very low energies.
In the paramagnetic state at high temperatures, the intensity of the signal gradually decreases again for increasing temperatures.

The seemingly smooth transition despite the first-order nature of the skyrmion-paramagnetic
phase boundary \cite{Bauer2013} could be taken as an indication of a mixed phase.
Such a mixed phase could either be understood as paramagnetic fluctuations being present in the skyrmion phase
or characteristics of the skyrmions surviving into the paramagnetic state.
As the elastic scans show that the paramagnetic state is clearly separated from the skyrmion phase by the onset of the smeared-out fluctuations
of the fluctuation-disordered paramagnetic phase, that cannot be observed in the skyrmion order, we rule out the first case.
The second case would be in line with a previous study that suggests
that skyrmion correlations exist in the fluctuation-disordered phase well
beyond the transition temperature \cite{Kindervater2019}.

The second result of our study shows that the non-reciprocity of the magnons in the skyrmion phase
is retained even for high temperatures inside the paramagnetic phase.
This result is due to the applied magnetic field.
The asymmetric energy shift diminishes with increasing temperature.
A similar result was also obtained for the very low-energy excitations in the $\mathrm{\upmu eV}$ region
that was measured by Soda \textit{et al.}.
They put forth the hypothesis that the paramagnetic fluctuations remain in a skyrmion-like state \cite{Soda2023}.
We believe in a simpler interpretation,
namely that the presence of an external field causes strong fluctuations
in the paramagnetic phase where long-range collective modes are still partially possible.
Apart from the external field, the persisting non-reciprocal character stems from the Dzyaloshinskii-Moriya interaction,
which originates from the crystal's symmetry alone.

No non-reciprocal characteristics were observed for the helimagnetic-paramagnetic transition at zero field,
where time-reversal symmetry is not broken.
Even though they were not detectable in the present experiment, minute polarisation-dependent asymmetries
in the chiral fluctuations still persists into the paramagnetic phase and
could still be observed by Roessli \textit{et al.} \cite{Roessli2002} using linear polarisation analysis.

Materials with persisting non-reciprocal responses well above the critical temperature may prove especially
interesting for research in the field of magnonics. A magnonic component such as a unidirectional field-guide would not need to
be cooled down to the onset of ordered magnetism, but could function at higher temperatures.

This work stets the stage for future investigations into the dynamics of the magnons at the
border of the conical and the skyrmion phase where we expect to observe similar transitional effects.

\appendix
\section{Unsubtracted Data}
\begin{figure*}[htb]
\begin{centering}
	\begin{tikzpicture}
		\draw (0, 0) node [ inner sep = 0 ] { \includegraphics[width=0.42\textwidth]{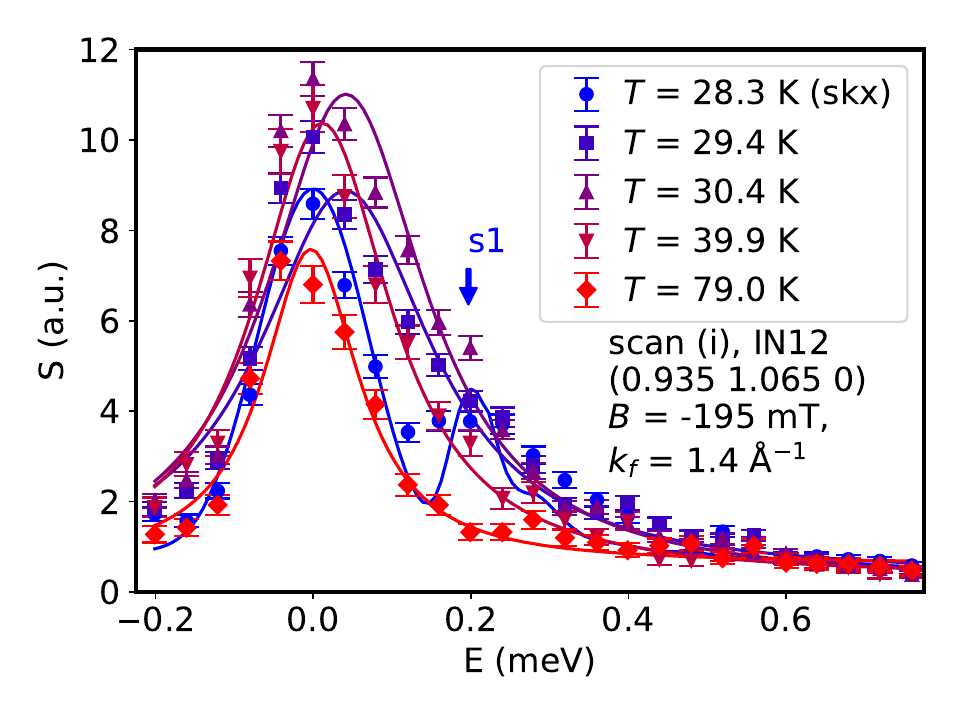} };
		\draw (-3.4, 2) node { \bf \fontfamily{phv} \selectfont (a) };
	\end{tikzpicture}
	\begin{tikzpicture}
		\draw (0, 0) node [ inner sep = 0 ] { \includegraphics[width=0.42\textwidth]{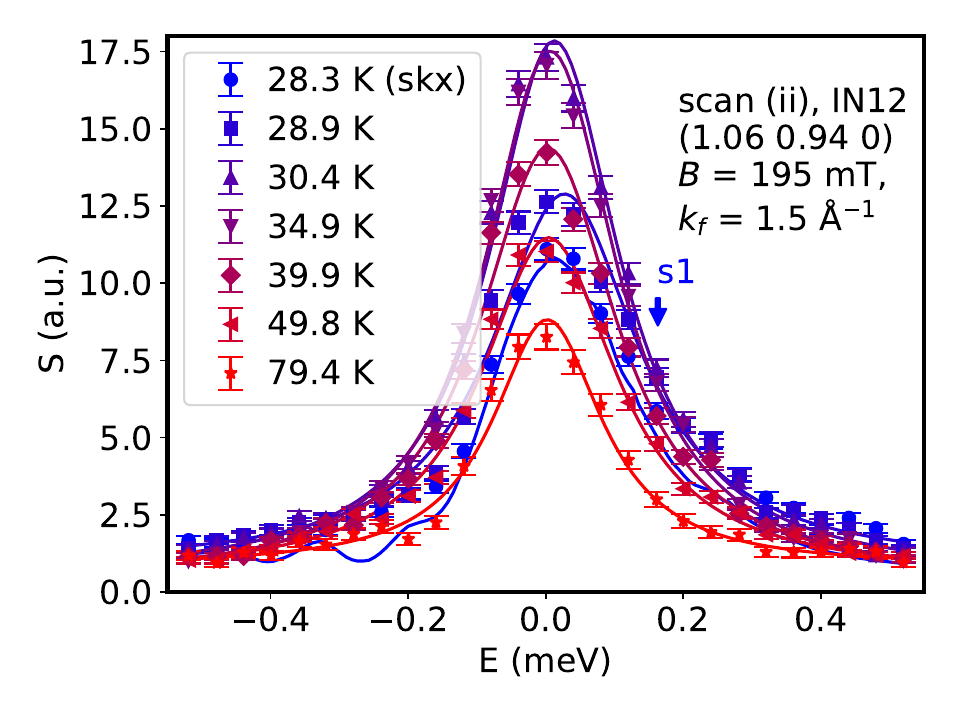} };
		\draw (-3.4, 2) node { \bf \fontfamily{phv} \selectfont (b) };
	\end{tikzpicture}
	\caption{Unsubtracted data sets with panels (a) and (b) corresponding to Fig. \ref{fig:IN12_skx_T} (c) and (d), respectively.
	Here, we also show the high-temperature non-magnetic reference scan at $T = 79\,\mathrm{K}$ explicitly.
	The solid lines for the skyrmion phase at $T_{skx}=28.3\,\mathrm{K}$ are resolution-convolution
	simulations of the magnon model \protect\cite{Garst2017, Weber2022Skyrmi} plus Gaussian profiles
	modeling the incoherent-elastic contribution, all the other solid lines are Lorentzian fits.}
	\label{fig:IN12_skx_T_nondiff}
\end{centering}
\end{figure*}

This appendix includes the raw, unsubtracted inelastic data sets for the skyrmion-paramagnetic transition.
They are shown in Fig. \ref{fig:IN12_skx_T_nondiff}.

\section*{Author Contributions}
\label{sec:data_appendix}
T.W. planned and performed the experiments and the data analysis and wrote the manuscript.
K.S. was responsible for the instrument.
A.B. and C.P. grew the crystal used in Sec \ref{sec:IN12_para}.
M.G. and J.W. created the theoretical skyrmion model.
All authors discussed the manuscript.

\begin{acknowledgements}
The measurements were conducted at the Collaborative Research Group (CRG) instrument \textit{IN12} \cite{IN12}
operated in collaboration with the CEA Grenoble at the ILL.
Crystal alignment was performed using the \textit{OrientExpress} \cite{OrientExpress} Laue diffractometer at the ILL.
The experiment has the DOI \href{https://doi.ill.fr/10.5291/ILL-DATA.CRG-3132}{10.5291/ILL-DATA.CRG-3132}.

We thank B. Vettard for technical assistance at IN12.
Thanks to P. B\"oni for providing the MnSi crystal used in Sec. \ref{sec:IN12_para_heli},
which has its origin at Tohoku University in Japan.
We furthermore thank M. Kugler for his original \textit{Python} implementation of the skyrmion model,
as well as D. Fobes and L. Beddrich for their independent \textit{Python} implementations.
\end{acknowledgements}

\section*{Funding}
This study was funded by the Deutsche Forschungsgemeinschaft (DFG, German Research Foundation) under TRR360 (Constrained Quantum Matter, Project No. 492547816),
SPP2137 (Skyrmionics, Project No. 403191981, Grant PF393/19), and the excellence cluster MCQST under Germany's Excellence Strategy EXC-2111 (Project No. 390814868).
Financial support by the European Research Council (ERC) through Advanced Grant No. 788031 (ExQuiSid) is gratefully acknowledged.

There are no conflicts of interest.

\section*{Data Availability}
The data files and the data analysis scripts are available under the DOI
\href{https://doi.org/10.5281/zenodo.17338264}{10.5281/zenodo.17338264}.
The source code for the model \cite{Weber2022Skyrmi} is available under the DOI
\href{https://doi.org/10.5281/zenodo.5718363}{10.5281/zenodo.5718363}.



%

\end{document}